\documentclass[opre,nonblindrev]{informs3} 

\usepackage{endnotes}
\let\footnote=\endnote

\usepackage{natbib}
 \bibpunct[, ]{(}{)}{,}{a}{}{,}%
 \def\bibfont{\small}%
 \def\BIBand{and}%

\TheoremsNumberedThrough     
\ECRepeatTheorems

\EquationsNumberedThrough    

\usepackage{epstopdf}
\usepackage{algorithm}
\usepackage{algorithmic}
\usepackage{latexsym}
\usepackage{url}
\usepackage{amssymb}

\def\C{\mathcal{C}}
\def\S{\mathcal{S}}
\def\HH{\mathcal{H}}
\def\pp{\mathbf{p}}
\def\PP{\mathbf{P}}
\def\Z{\mathcal{Z}}
\def\F{\mathcal{F}}
\def\X{\mathcal{X}}

\def\E{\mathbb{E}}
\def\P{\mathbb{P}}
\def\R{\mathbb{R}}
\def\G{\mathcal{G}}
\def\N{\mathbb{N}}

\def\p{\mathbf{\pi}}

\def\1{\mathbf{1}}
\def\maketitle{%
  \newpage
  \theARTICLETITLE
  \theARTICLEAUTHORS
  \theARTICLEABSTRACT
  \theARTICLERULE
}

\def\setoddRF{\hfil} \def\setevenRF{\hfil} 

\def\setoddRH{\hbox to \textwidth{\fs.7.8.\tabcolsep0pt
  \begin{tabular*}{\textwidth}[b]{l@{\extracolsep\fill}r}
  &\raisebox{0pt}[0pt][0pt]{\fs.10.10.\thepage}\\[-4pt]
  \rlap{\VRHDW{0.5pt}{0pt}{\textwidth}}&\\
  \end{tabular*}}}
\def\setevenRH{\hbox to \textwidth{\fs.7.8.\tabcolsep0pt
  \begin{tabular*}{\textwidth}[b]{l@{\extracolsep\fill}r} \raisebox{0pt}[0pt][0pt]{\fs.10.10.\thepage}\\[-4pt]
  \rlap{\VRHDW{0.5pt}{0pt}{\textwidth}}&\\
  \end{tabular*}}}
  
\begin{document}


\RUNAUTHOR{Kazerouni and Van Roy}

\RUNTITLE{Learning to Price with Reference Effects}

\TITLE{Learning to Price with Reference Effects}

\ARTICLEAUTHORS{%
\AUTHOR{Abbas Kazerouni}
\AFF{Department of Electrical Engineering, Stanford University, Stanford, CA 94305, \EMAIL{abbask@stanford.edu}} 
\AUTHOR{Benjamin Van Roy}
\AFF{Departments of Management Science and Engineering and Electrical Engineering, Stanford University, CA 94305, \EMAIL{bvr@stanford.edu}}
} 

\ABSTRACT{%
As a firm varies the price of a product, consumers exhibit reference effects, making purchase decisions based not only on the prevailing price but also the product's price history.  
We consider the problem of learning such behavioral patterns as a monopolist releases, markets, and prices products.  This context calls for pricing decisions that intelligently trade off between maximizing revenue generated by a current product and probing to gain information for future benefit.  Due to dependence on price history, realized demand can reflect delayed consequences of earlier pricing decisions.  As such, inference entails attribution of outcomes to prior decisions and effective exploration requires planning price sequences that yield informative future outcomes.  Despite the considerable complexity of this problem, we offer a tractable systematic approach.  In particular, we frame the problem as one of reinforcement learning and leverage Thompson sampling.  We also establish a regret bound that provides graceful guarantees on how performance improves as data is gathered and how this depends on the complexity of the demand model.  We illustrate merits of the approach through simulations.
}%



\maketitle

%


\section{Introduction}

Consider a monopolist that prices and sells a variety of products over time.  Accounting for the impact of prices on demand can greatly improve revenue.  
This dependence is often complex, with purchase decisions influenced not only by prevailing prices but also price histories.  For example, 
purchases can be triggered by price reductions or prices of alternatives.  In this paper we develop an approach to learning such behavioral patterns
through setting prices and observing sales, with a goal of maximizing cumulative revenue over the course of many product life cycles.

Efficient learning calls for a thoughtful balance between maximizing revenue generated by a current product and probing to gain information 
that can be leveraged to increase subsequent revenue.  Reference effects, by which we mean dependencies of current demand on past prices,
bring substantial complexity to this so-called {\it exploration-exploitation dilemma}.  First, there can be ambiguity as to whether purchases are
triggered by the current price or some relation to past prices.  To disambiguate, an effective learning algorithm must attribute delayed consequences 
to inter-temporal pricing decisions.  Second, exploration entails coordinated selection of price sequences; independent selection 
of spot prices may not suffice.  This is because demand may respond favorably to particular price histories, and probing appropriately selected
price sequences can be required to efficiently learn that.

Despite the considerable complexity of this problem, we provide what to our knowledge is the first tractable systematic approach.
We proceed by framing the problem as one of reinforcement learning and then, for particular classes of demand models, developing 
a computationally efficient learning algorithm based on Thompson sampling.  To offer some assurance of statistical efficiency, 
we establish a bound on expected regret.  With respect to the number of past products $K$,  the bound grows as 
$\sqrt{K} \log K$, which indicates that per-period expected regret vanishes over time.
The bound applies very broadly across model classes, as it depends on the Kolmogorov and
eluder dimensions, which are statistics that quantify model complexity in relation to data requirements for
model fitting and for exploration, respectively.  In particular, the dominant term in our bound grows with the geometric mean
of the two notions of dimension.  We also present simulation results that demonstrate strong performance relative
to less sophisticated exploration schemes.

For the sake of exposition, most of our discussion will focus on a simplified setting in which the firm sells indistinguishable products 
in sequence under unchanging market conditions, discontinuing each product before launching the next, and with each product 
marketed over a fixed number $H$ of time periods.  The price can be adjusted in each of these time periods,
and the demand for a product in any given period depends on its prevailing and previous prices.  In Section \ref{generalSec},
we explain how algorithms and results can be extended to treat more complex models that capture important
features of realistic problems.  This includes models with covariates that capture distinguishing features of products 
and varying market conditions and that allow for simultaneous pricing and sales of multiple products with 
overlapping life cycles of varying duration.

There is a substantial literature on pricing with reference effects.  \cite{mazumdar2005reference} provides a comprehensive survey that covers 
both behavioral research that provides evidence and examines the structure of reference effects and methodological research on how
pricing strategies should respond.  Strategies for particular model classes have been developed in 
\cite{greenleaf1995impact, kopalle1996asymmetric, 
 fibich2003explicit, ahn2007pricing,popescu2007dynamic, heidhues2014regular}. 
However, these papers treat the problem of pricing given known demand models, with no learning required.
Electronically-mediated markets, the increasing availability of data, and advances in the field of 
machine learning have fueled a vast and growing literature on learning to price.
We refer the reader to \cite{den2015dynamic} for a comprehensive review of the literature and research directions. 
To our knowledge, our work is the first to treat learning in the presence of reference effects.  

The fact that pricing decisions can result in delayed consequences has presented an obstacle to the development of efficient algorithms that 
learn to price effectively with reference effects.  In this paper, we offer a new approach 
through framing the problem as one of reinforcement learning and bringing to bear recent developments in the application of Thompson sampling 
to such problems.  It is worth noting, however, that we are not the first to apply Thompson sampling to a pricing problem.  In particular, \cite{ferreira2015online} 
considers an approach based on Thompson sampling to address a multiproduct pricing problem with resource constraints, though without reference effects.

Our pricing strategy is based on the  posterior sampling reinforcement learning (PSRL) algorithm, originally proposed by Strens \cite{strens2000bayesian} under the name ÒBayesian Dynamic Programming,Ó as a heuristic for reinforcement learning in Markov decision processes.  Building on general results established in \cite{dan1} for Thompson sampling, regret analyses for PSRL are developed in \cite{ian1,osband2014model}.  In principle, results of \cite{osband2014model} apply to the problem we consider in this paper, but the associated regret bound depends on a Lipschitz constant which is not clear how to characterize in our context.  We instead build directly on the technical tools of \cite{dan1} and \cite{osband2014model} to derive custom regret bounds for PSRL in our pricing model.

The rest of this paper is organized as follows. In Section \ref{setupsec}, we formulate a dynamic pricing problem that addresses reference effects.  In Section \ref{TPSec} , we propose Thompson Pricing (TP) as a heuristic strategy for the problem and provide a general regret bound in Theorem \ref{mainth} of that section.  To carry out a more concrete study, in Section \ref{specialSec}, we consider the special case of linear demand.  For this context, we specialize and interpret regret bounds and present computational results that illustrate merits of TP.  Section \ref{generalSec} discusses how our dynamic pricing model and algorithm can be generalized to accommodate complexities arising in practical settings, such as observation of covariates that inform demand forecasts, coordinated pricing across multiple products, and pricing of products with overlapping sales seasons.  Section \ref{analysissection} presents an analysis of TP in a general setting that accommodates various aforementioned complexities, leading to the main technical result of the paper (Theorem \ref{mainth0}). We offer concluding remarks in Section \ref{ConcSec}.

\section{Problem Formulation}
\label{setupsec}

In this section, we formulate a dynamic pricing problem and highlight the role played by reference effects.  To facilitate exposition of core ideas that we will develop in the paper, our model leaves out many  complexities that may be required to adequately address practical contexts.  In Section \ref{generalSec}, we will discuss how the model and ideas can be extended to accommodate some such complexities.

Consider a monopolist selling indistinguishable products over a sequence of sales seasons under unchanging market conditions.  We will think of each sales season as an {\it episode} of interaction between the monopolist and the consumer market.  Let each episode last for $H$ time periods.  At the start of each time period, the monopolist sets a price, observes random demand, and collects revenue. We assume that the monopolist faces no supply constraint, so that all demands are met.  As an illustration, one might think of the monopolist as a seller of coats that are sold over the Autumn and Winter who adjusts price over each week.  In this case, each episode is a six month period and each period lasts a week.

At the start of each period $h$ of an episode $k$, the seller sets a price $p_{k,h}$ and observes demand $y_{k,h}$, which,  conditioned on information available when the price is set, is log-normally distributed with parameters $d_{k,h}-{\sigma^2}/{2}$ and $\sigma^2$.  Hence, $d_{k,h}$ denotes expected demand, while $\sigma^2$ represents uncertainty.  The expected revenue upon setting the price is given by $r_{k,h} = d_{k,h} p_{k,h}$.

The expected demand for a product may depend on factors such as the quality of the item, the price, and consumer behavior. 
If the monopolist does not understand these dependencies {\it a priori}, in order to identify an optimal price, he must learn through experimentation.

\subsection{Memoryless Demand}

In the simplest case, one might assume that expected demand depends only on the current price; that is $d_{k,h} = f_\theta(p_{k,h})$ for some function $f_\theta:\R^+\to \R$, where $\theta \in \R^l$ is an unknown parameter representing what the monopolist does not know about demand structure.  Given knowledge of $\theta$, the monopolist should maintain a constant price $p = \arg\max_p p f_\theta(p)$ over time.
However, a monopolist may have to learn $\theta$ through experimentation.  This calls for a pricing strategy that balances between exploration and exploitation and converges over time to an optimal price.  This can be viewed as a structured bandit learning problem, and several pricing algorithms have been proposed for variations of the problem \cite{ferreira2015online,kincaid1963inventory,gallego1994optimal,gallego1997multiproduct,
 bitran1998coordinating,besbes2009dynamic,araman2009dynamic, farias2010dynamic,besbes2012blind,lobo2003pricing}.   
  
In the memoryless demand model we have described, the way in which consumers respond to a current price does not depend on price history.  In reality, reference effects play a substantial role in purchase decisions \cite{mazumdar2005reference}.  For example, offering a discount often increases demand not only because the new price is low, but also because it is lower than the previous price.  Black Friday and Cyber Monday sales constitute well-known examples of this phenomenon.  While such reference effects naturally occur they cannot be learned using memoryless demand models.  A broader class of dynamic models is called for, as well as more sophisticated strategies involving strategic sequencing of prices to maximize cumulative revenue.  We now discuss such a model.

\subsection{Reference Effects}
\label{stsetup}

We consider a model in which expected demand over a time period may depend not only on current price but also on $n$ previous prices within the current episode.  Here, $n$ is a parameter that represents duration of memory in the demand model.  In any period $h = 1,\ldots,H$ of episode $k$, we consider the state of the demand model to be 
\begin{equation}
\label{statedef}
s_{k,h} = [p_{k,{\max(1,h-n)}},\cdots,p_{k,{h-2}},p_{k,{h-1}}]^\top,
\end{equation}
which represents the $n$-step price history of the product. 
Then, the expected demand at period $h$ is taken to be 
\begin{equation}
\label{expdemstrategic}
d_{k,h} = f_\theta(p_{k,h},s_{k,h}),
\end{equation}
for a demand function $f_\theta$, which depends on an unknown parameter $\theta$.  Dependence of the expected demand on the state $s_{k,h}$ captures reference effects.

Note that prices in environments with reference effects bear delayed consequences: a price does not only influence immediate but also future demand.  Therefore, given knowledge of $\theta$, an optimal pricing strategy does not fix a constant price, as would be the case with a memoryless demand model, but rather, plans a sequence of prices that vary over periods of the episode.  Such a sequence $\pp = (p_1,p_2,\cdots,p_H)$ generates expected revenue
\begin{equation}
\label{valfunc}
V_\pp = \sum_{h=1}^H p_h f_{\theta}(p_h,s_h),
\end{equation}
over the episode, where $s_h$ is the state at the start of period $h$, under this price sequence.
Therefore, an optimal price sequence is given by
\begin{equation}
\label{optpol}
\pp^* = \argmax_{\pp}{~V_\pp}.
\end{equation}
Note that this optimization problem can be solved via dynamic programming. We write 
\begin{equation}
\label{optpol2}
\pp^* = \text{DP}(\theta,H,p_{\max})
\end{equation}
to indicate that $\pp^*$ is the solution of the associated dynamic program applied to above optimization problem with  environment variable $\theta$ and horizon $H$.  We also provide an argument for a price constraint $p_{\max}$; if $p_{\max} < \infty$ then each price is constrained to the interval $[0,p_{\max}]$.

Since the environment is unknown to the seller, he must experiment to learn the demand model while earning revenue.  A pricing strategy ${S}$ is a sequence of policies such as $\left(\pp_1,\pp_2,\pp_3,\cdots\right)$ to be executed in consecutive episodes, where  policy $\pp_k$ is a (possibly random) function of the history observed prior to episode $k$.  We will assess the performance of a  pricing strategy $S$ in terms of its cumulative regret over $K$ episodes  defined by
\begin{equation}
\label{Regretdef}
R_K(S) = \E\left[\sum_{k=1}^K \left(V_{\pp^*} - V_{\pp_k}\right)\right].
\end{equation}

With reference effects, the agent has the ability to influence purchasing behavior by exploiting the manner in which consumers react to price trajectories.  As such, the learning problem involves learning how to influence consumer behavior, which is a deeper issue than that addressed when learning with a memoryless demand model.

\section{Thompson Pricing}
\label{TPSec}

In this section, we present Thompson Pricing (TP), a heuristic strategy that learns to price with reference effects and bound its cumulative regret.

\subsection{The Algorithm}
With TP, the seller begins with a prior distribution $\p$ over the unknown parameter $\theta$.  Let 
$$\HH_k = \left\{\left(p_{j,1},s_{j,1},y_{j,1},\cdots,p_{j,H},s_{j,H},y_{j,H}\right) : j =1,\ldots,k-1\right\}$$
denote the history of observations made prior to the start of the $k$th episode.  Based on this history, the agent generates a sample $\hat\theta_{k}$ from the posterior distribution $\p\left(\cdot|\HH_k\right)$ and then, treating this sample as truth, computes the policy 
\begin{equation}
\label{sampol}
\pp_{k} = \text{DP}(\hat\theta_{k},H,p_{\max}).
\end{equation}
The resulting policy is applied through  episode $k$. 
After episode $k$, the posterior distribution is updated based on observations made over the episode, and the process repeats. 
A more precise description of TP is provided as Algorithm \ref{alg1}.

\begin{algorithm}[t]
   \caption{Thompson Pricing  (\texttt{TP})}
   \label{alg1}
\begin{algorithmic}
   \STATE {\bfseries Input:} episode length $H$, function $f$, maximum price $p_{\max}$ and prior $\p$
   \STATE{\bfseries Initialize:} $\HH_1 = \emptyset$
   \FOR{$k=1,2,\cdots$}   
   \STATE Sample $\hat \theta_{k} \sim \p(\cdot|\HH_k)$
   \STATE Compute $\hat\pp_k = (p_{k,1},p_{k,2},\cdots,p_{k,H}) = \text{DP}(\hat\theta_k,H,p_{\max})$
   \FOR{$ h=1,2,\cdots,H$}
   		\STATE Set price $p_{k,h}$
   		\STATE Observe random demand $y_{k,h}$
   \ENDFOR
   \STATE Update $\HH_{k+1} = \HH_k\cup\{\left(p_{k,1},s_{k,1},y_{k,1},\cdots,p_{k,H},s_{k,H},y_{k,H}\right)\}$
   \ENDFOR
   
\end{algorithmic}
\end{algorithm} 

Note that TP determines the price trajectory for the entire episode at the start of each episode.  Each trajectory can probe the market to reveal consequences 
not only of individual prices but of a price sequence.  Sampling from the posterior distribution over models trades off between exploiting what has been learned
and exploring the unknown.

\subsection{Regret Bound}

In principle, TP can be applied with any distribution over demand functions, though computational requirements vary greatly depending on the problem class.
We now provide a general regret bound that applies broadly.  In subsequent sections, we specialize TP and its regret bound to more specific problem classes
that admit efficient computation.

Let $\Theta$ denote the support of the prior distribution $\p$.  
In each time period, the state of the system is characterized by prices quoted so far within the episode.  As such, the state space is given by
$\S = \emptyset\cup_{i=1}^n [0,p_{\max}]^i$.  The demand function $f_\theta$ maps the current price and state to expected demand.  Hence, the set of possible
demand functions is given by
\begin{equation}
\label{Fdef}
\F = \left\{f_\theta:[0,p_{\max}]\times \S\to\R\big|\theta\in\Theta\right\}.
\end{equation}
We assume that the range of the demand function is bounded.
\begin{assumption}
\label{revassump}
There exists $d_{\max}>0$ such that, for all $\theta \in \Theta$, $p \in [0,p_{\max}]$, and $s \in \mathcal{S}$, $f_\theta(p,s) \in [0, d_{\max}]$.
\end{assumption}

We will provide a regret bound that applies to any class of demand functions. The dependence of regret on the class of demand functions can be characterized by statistics that reflect  suitable notions of complexity. The regret bound that we will provide depends on two such statistics. 

Let $\G$ be a collection of functions, each mapping a set $\X$ to $\R$. For all $\alpha>0$, let $N(\G,\alpha)$ denote the $\alpha$-covering number of $\G$ with respect to the supremum norm. 
\begin{definition}
Let $\G$ be a collection of functions, each mapping a set $\X$ to $\R$. The Kolmogorov dimension of $\G$, denoted by $d_K(\G)$, is
$$d_k(\G) = \limsup_{\alpha\downarrow 0}\frac{\log N(\G,\alpha)}{\log(1/\alpha)}.$$
\end{definition}
The Kolmogorov dimension is a notion of complexity commonly used to quantify the number of data samples required to avoid statistical overfit.  Sample complexity results in statistical learning
that build on this concept typically apply to contexts where data is generated by a stationary source.  In our pricing problem, data is not produced by an exogenous stationary source
but rather through probing actions that hone in on an optimal price sequence.  To bound sample requirements in such a context, a new notion of complexity is called for.  To serve this need, 
we will use the {\em eluder dimension}, as introduced in \cite{dan1}.  To define this, we begin with a notion of dependence.
\begin{definition}
Let $\G$ be a collection of functions, each mapping a set $\X$ to $\R$.  An element $x \in \X$ is $\epsilon$-{\it dependent} on $\{x_1,\ldots,x_l\} \subseteq \X$ with respect to $\G$ if any pair of functions $g_1,g_2$ satisfying $\sqrt{\sum_{j=1}^{l}\left(g_1(x_j)-g_2(x_j)\right)^2} \leq \epsilon$ also satisfies $g_1(x) - g_2(x) \leq \epsilon$.  Further, $x$ is $\epsilon$-{\it independent} of $\{x_1,\ldots,x_l\}$ with respect to $\G$ if $x$ is not $\epsilon$-dependent on $\{x_1,\ldots,x_l\}$.
\end{definition}
Intuitively, an action $x$ is independent of $\{x_1,\ldots, x_l\}$ if two functions that make similar predictions at $\{x_1,\ldots,x_l\}$ can nevertheless differ significantly in their predictions
at $x$.  This concept suggests the following notion of dimension.
\begin{definition}
\label{eldim}
Let $\G$ be a collection of functions, each mapping a set $\X$ to $\R$.  The $\epsilon$-eluder dimension $d_E(\G,\epsilon)$ is the length $l$ of the longest sequence
of elements of $\G$ such that, for some $\epsilon' \geq \epsilon$, every element is $\epsilon'$-independent of its predecessors.
\end{definition} 

The following theorem provides a general regret bound for TP. 
\begin{theorem}
\label{mainth}
Consider dynamic pricing with reference effects, as formulated Section \ref{stsetup}. Under Assumption \ref{revassump},
\begin{equation}
\label{regretbound}
R_K(TP) \leq p_{\max}\left(1 + Hd_{\max} d_E(\F,(KH)^{-2})+4\sqrt{\beta_Kd_E(\F,(KH)^{-2})KH}\right) +\frac{4p_{\max}d_{\max}}{KH},
\end{equation}
where 
\begin{equation}
\label{maintheq2}
\beta_K = 8\sigma^2\log((KH)^2 N(\F,(KH)^{-2})) + \frac{2}{KH}\left(8d_{\max}+\sqrt{8\sigma^2\log 4}\right),
\end{equation}
and, in an asymptotic notation, 
\begin{equation}
\label{maintheq3}
R_K(TP) = O\left(p_{\max}\sigma\sqrt{d_K(\F)d_E(\F,(KH)^{-2})KH\log(KH)}\right).
\end{equation}
\end{theorem}
Theorem \ref{mainth} provides an upper bound on the regret of TP applied to an arbitrary class of demand functions.
The regret bound established in this theorem depends on the geometric mean of Kolmogorov and eluder dimensions. Furthermore, this regret bound is increasing in the maximum price $p_{\max}$ and demand uncertainty $\sigma$.
The proof of this theorem is quite involved and presented in Section \ref{analysissection}. In fact, the analysis of Section \ref{analysissection} addresses a more general problem that allows for covariates and multiproduct pricing.  Theorem \ref{mainth} follows immediately from the more general result.

\section{A Linear Demand Model}
\label{specialSec}

In this section, we study the special case of a linear demand model, in which
\begin{equation}
\label{expdemstlin}
f_\theta(p_{k,h},s_{k,h}) = \left\{ \begin{array}{ll}
\alpha +  \beta p_{k,h}	\qquad  &\text{if }h=1\\
\alpha +  \beta p_{k,h}+ \phi_{h-1}^\top s_{k,h} \qquad
&\text{if }2\leq h\leq n\\
\alpha + \beta p_{k,h}+ \phi_{n}^\top s_{k,h}	 \qquad
&\text{if }n+1\leq h
\end{array}
\right .
\end{equation}
where $\alpha,\beta\in\R$, $\phi_1 \in \R^1, \ldots, \phi_n \in \R^n$ are unknown parameters of the demand function. 
Note that the state vector increases in length over the first $n$ time periods, and $\phi_1,\ldots, \phi_n$ represent coefficient vectors that 
multiply state vectors of different lengths.  The model includes a total of $2 + n(n+1)/2$ unknown parameters, which can be encoded in terms of a vector 
$\theta = \begin{bmatrix} \alpha,\beta , \phi_1^\top,\cdots, \phi_n^\top\end{bmatrix}^\top$.
We consider a normal prior distribution over $\theta$ with mean $\mu$ and covariance matrix $\Sigma$.

Thanks to conjugacy properties of normal distributions, the posterior distribution of $\theta$ after any number of episodes remains normal. 
To specify the update rules for posterior means and covariances, let us define a few auxiliary variables.
Given observations  $\left( p_{k,1},s_{k,1},y_{k,1},\cdots,p_{k,H},s_{k,H},y_{k,H}\right)$ gathered over the $k$'th episode, define
$$
w_{k,h} = \log(y_{k,h}) + \frac{\sigma^2}{2},$$
and for $h = 1,\ldots, H$, let
$$x_{k,h} = \left\{ \begin{array}{ll} 
 \big[1,p_{k,h}  , \overbrace{0\cdots,0}^{\frac{ n(n+1)}{2}}\big] \qquad & \text{if  } h=1\\
 \big[1,p_{k,h} ,\overbrace{0\cdots,0}^{\frac{(h-1)(h-2)}{2}}, s_{k,h}^\top, \overbrace{0\cdots,0}^{\frac{n(n+1)-h(h-1)}{2}}\big] \qquad & \text{if  } 2\leq h\leq n\\
\big[1,p_{k,h},\overbrace{0\cdots,0}^{\frac{n(n-1)}{2}}, s_{k,h}^\top\big] \qquad & \text{if  } h\geq n+1 . 
\end{array}\right .$$
Then, the mean and covariance matrix of the posterior distribution are updated according to 
\begin{equation}
\label{update2}
\mu\leftarrow \left(\Sigma^{-1} + \frac{1}{\sigma^2}\sum_{h=1}^H x_{k,h}^\top x_{k,h}\right)^{-1}\left(\Sigma^{-1}\mu + \frac{1}{\sigma^2}\sum_{h=1}^H w_{k,h} x_{k,h}^\top\right),~~~~\Sigma\leftarrow \left(\Sigma^{-1} + \frac{1}{\sigma^2}\sum_{h=1}^H x_{k,h}^\top x_{k,h}\right)^{-1}.
\end{equation}

At the start of each $k$'th episode, TP samples $\hat\theta_k$ from the prevailing posterior distribution and applies the policy
\begin{equation}
\label{pklindem}
\pp_k = \mbox{DP}(\hat\theta_k,H,p_{\max})
\end{equation} 
throughout the episode.  Note that the state evolves deterministically over the episode; that is, the state in any time period is determined by the state in the previous time period and the selected price.
With these linear dynamics and linear demand model, the dynamic program reduces to a quadratic optimization problem which can be solved efficiently, as 
carried out by Algorithm \ref{alg3}. 

To illustrate successive steps of Algorithm \ref{alg3}, we introduce some notation. Given an $l_1\times l_2$ matrix $M$ and for $1\leq i\leq l_1,1\leq j\leq l_2$, we let $M[i,j]$ denote the $(i,j)$'th element of $M$. For $1\leq i_1<i_2\leq l_1$, we take $M[i_1:i_2,j]$ to be the submatrix of $M$ consisting of the elements in column $j$ and rows $i_1$ to $i_2$. Similarly for $1\leq j_1<j_2\leq l_2$, $M[i,j_1:j_2]$ denotes the submatrix of $M$ consisting of the elements in row $i$ and columns $j_1$ to $j_2$. 
By \eqref{valfunc} and \eqref{expdemstlin}, the expected revenue of  policy $\pp = (p_1,p_2,\cdots,p_H)\in [0,p_{\max}]^H$ in an episode is
\begin{equation}
\label{valuelindemand}
V_\pp = \sum_{h=1}^H \alpha p_h + \sum_{h=1}^H \beta p_h^2 + \sum_{h=2}^n \alpha p_h \phi_{h-1}^\top s_h + \sum_{h=n+1}^H \alpha p_h  \phi_{n}^\top s_h,
\end{equation}
where $s_h = [p_{\max(1,h-n)},\cdots,p_{h-2},p_{h-1}]^\top$, for $2\leq h\leq H$. Now, let $M_\theta$ be an $H\times H$ matrix which satisfies 
\begin{enumerate}
\item $M_\theta[h,h] = \beta$ for $1\leq h\leq H$,
\item $M_\theta[\max(h-n,1):h-1,h] = \frac{1}{2}\phi_{\max(h-1,n)}$ for all $2\leq h\leq H$,
   \item $M_\theta[h,\max(h-n,1):h-1] = \frac{1}{2}\phi_{\max(h-1,n)}^\top$ for all $2\leq h\leq H$,
\item all other entries of $M$ are equal to 0.
\end{enumerate}
Given $M_\theta$, \eqref{valuelindemand} can be expressed as 
\begin{equation}
\label{valuelindemand2}
V_\pp = \pp^\top M_\theta \pp + \alpha \1^\top \pp,
\end{equation}
where with a slight abuse of notation, $\pp$ is treated as an $H$ dimensional vector. Therefore, given the sampled parameter vector $\hat\theta_k$, the dynamic program in \eqref{pklindem} is equivalent to 
\begin{equation}
\label{pklindem2}
\pp_k = \argmax_{\pp\in[0,p_{\max}]^H}{~\pp^\top M_{\hat\theta_k} \pp + \hat\alpha_k \1^\top \pp}.
\end{equation}
The quadratic optimization problem in \eqref{pklindem2} can be solved efficiently via the standard convex optimization tools provided that the matrix 
$M_{\hat\theta_k}$ is negative semi-definite. On the other hand, if $M_{\hat\theta_k}$ is not negative semi-definite, the optimization problem in \eqref{pklindem2} is NP-hard.
While, $M_{\hat\theta_k}$ is not guaranteed to be negative semi-definite for all realizations of $\hat\theta_k$, for appropriate values of the prior mean $\mu$ (for example with mean of $\beta$ being negatively large), $M_{\hat\theta_k}$ would be negative semi-definite with high probability. From a practical point of view, at the start of episode $k$, sampling from the posterior distribution can be repeated until the sampled $\hat\theta_k$ results in a negative semi-definite $M_{\hat\theta_k}$.
 Algorithm \ref{alg4} describes the specialization of TP to the described linear environment.

\begin{algorithm}[t]
   \caption{\texttt{DP-lin}}
   \label{alg3}
   \begin{algorithmic}
   \STATE {\bfseries Input:} $\theta,H,p_{\max}$
   \STATE {\bfseries Output:} $\pp$
   \STATE Extract $\alpha,\beta,\phi_1,\cdots,\phi_n$ from $\theta$
   \STATE{\bfseries Initialize:} $M = \mathbf{0}_{H\times H}$
   \STATE Set $M[1,1] = \beta$
   \FOR{$h=2,3,\cdots,H$}   
   \STATE Set $M[h,h]=\beta$
   \STATE Set $M[\max(h-n,1):h-1,h] = \frac{1}{2}\phi_{\max(h-1,n)}$
   \STATE Set $M[h,\max(h-n,1):h-1] = \frac{1}{2}\phi_{\max(h-1,n)}^\top $
   \ENDFOR
   \STATE Return $\pp=\argmax_{x\in[0,p_{\max}]^H}{~
x^\top Mx + \alpha\mathbf{1}^\top x}$
\end{algorithmic}
\end{algorithm} 

\begin{algorithm}[t!]
   \caption{\texttt{TP-lin}}
   \label{alg4}
\begin{algorithmic}
   \STATE {\bfseries Input:} $H,p_{\max},\mu_0,\Sigma_0,\sigma^2$
   \STATE{\bfseries Initialize:} $\mu=\mu_0,\Sigma=\Sigma_0$
   \FOR{$k=1,2,\cdots$}   
   \STATE Sample $\hat\theta_k\sim N(\mu,\Sigma)$
   \STATE Compute $\hat\pp_k=(p_{k,1},p_{k,2},\cdots,p_{k,H})=\texttt{DP-lin}( \hat\theta_k,H,p_{\max})$
   \FOR{$h=1,2,\cdots,H$}
   \STATE Set price $p_{k,h}$
   \STATE Observe random demand $y_{k,h}$
   \ENDFOR
   \STATE Update $\mu$ and $\Sigma$ according to \eqref{update2}
   \ENDFOR
\end{algorithmic}
\end{algorithm} 

In addition to TP, let us consider two other pricing strategies. First, consider a seller who is agnostic to reference effects and adopts a simple pricing strategy based on Thompson sampling   which is suitable for  memoryless demands. Specifically, such a seller assumes that the expected demand at period $h$ in episode $k$  is 
$$d_{k,h} = \beta p_{k,h} + \alpha,$$
for some unknown $\alpha,\beta\in\R$  and considers a normal prior distribution over $[\alpha, \beta ]^\top$ with mean $\mu$ and covariance matrix $\Sigma$. At the beginning of each period $h$ in episode $k$, $[\hat\alpha_{k,h}, \hat\beta_{k,h} ]^\top$ is sampled from the prevailing posterior distribution and the price 
$$p_{k,h} = \argmax_{p\in[0,p_{\max}]}{~ \hat\beta_{k,h} p^2 + \hat\alpha_{k,h} p}$$
is set for the product throughout the period. Upon observing the random demand $y_{k,h}$ at the end of this period, the posterior distribution remains normal with its mean and covariance  updated via 
$$\mu\leftarrow \left(\Sigma^{-1} + \frac{1}{\sigma^2} x_{k,h}x_{k,h}^\top\right)^{-1}\left(\Sigma^{-1}\mu + w_{k,h}x_{k,h}\right),~~\Sigma\leftarrow \left(\Sigma^{-1} + \frac{1}{\sigma^2} x_{k,h}x_{k,h}^\top\right)^{-1},$$
where  $x_{k,h} = [1,p_{k,h}]^\top$. 

The above memoryless pricing strategy, which performs near optimally in memoryless environments \cite{ferreira2015online}, will drastically fail  in the presence of reference effects. This failure can be attributed to ignorance towards the reference effects. With the above misspecified demand model, the seller is not taking the delayed consequences of prices into account while the optimal pricing strategy takes advantage this phenomenon. As a result, the memoryless pricing strategy is not able to learn the optimal pricing strategy in an environment with reference effects. 

As a second pricing strategy,  consider a seller who assumes the demand model of \eqref{expdemstlin}, but instead of TP employs a weak version of Thompson sampling as follows. At the beginning of each period within an episode, a model is sampled from the prevailing  posterior distribution and treating it as  truth, a price is set greedily to maximize the expected immediate revenue. Upon observing the random demand at the end of the period, the posterior parameters are updated according to \eqref{update2} and the process repeats. 

In this pricing strategy, the seller indeed accounts for the effect of previous prices on current demand when maximizing the expected immediate revenue, but he overlooks the effect of current price on future demands. This is while the optimal pricing strategy determines the price trajectory for an  episode in a way to fully exploit the delayed consequences of the prices and maximize the total expected revenue. For example, the optimal pricing strategy might suggest  keeping the prices low (and collect a low revenue) at the initial few periods in exchange for large demands (and large revenues) at the subsequent periods. The greedy behavior of the weak version of Thompson sampling does not allow for such strategic plannings and hence prevents the seller from learning the optimal strategy. 

\begin{figure}[t]
\centering
  \includegraphics[width=0.6\linewidth]{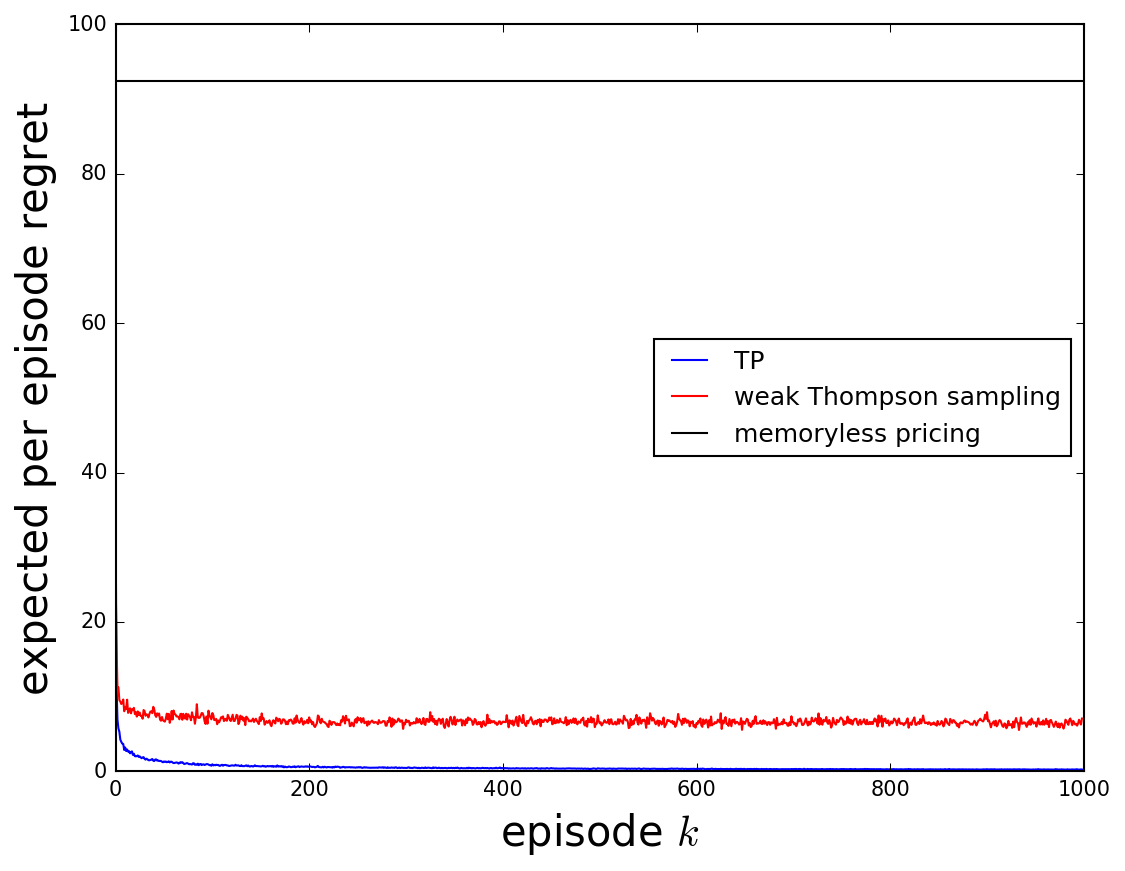}
\caption{Average per episode regret of TP, memoryless pricing and weak Thompson sampling in a linear regime with reference effects.}
\label{fig1}
\end{figure}

To compare the performance of TP with the above two alternative pricing strategies, we simulate an environment with linear demand as described. 
 In the simulation, we let $H=20$,  $n = 6$, $p_{\max}=1$ and $\sigma^2=10$. We assume that the prior distributions of $\alpha$ and $\beta$ are $N(7.5,10)$ and $N(-4,10)$, respectively. Further, we assume that each component of $\phi_i$ has a $N(0,10)$ prior distribution for $1\leq i\leq n$.  Figure \ref{fig1} shows the  per-episode regret of these three pricing strategies which are  averaged over thousand random realizations. As depicted in this figure, TP (Algorithm \ref{alg4}) quickly learns the optimal strategy and hence its per-episode regret diminishes quickly. However, as a result of  model misspecification, the memoryless pricing which ignores the reference effects fails to learn the unknown parameters and  suffers from a large non-diminishing per-episode regret. This observation points out that more sophisticated pricing strategy is required in the presence of reference effects and neglecting such effects massively degrades the performance.
Figure \ref{fig1} also depicts the per-episode regret achieved by the weak version of Thompson sampling described above. As discussed earlier, this version of Thompson sampling does not plan for the future  and, as Figure \ref{fig1} shows, its per-episode regret converges to a non-zero constant. 

\begin{figure}[t]
\centering
  \includegraphics[width=0.6\linewidth]{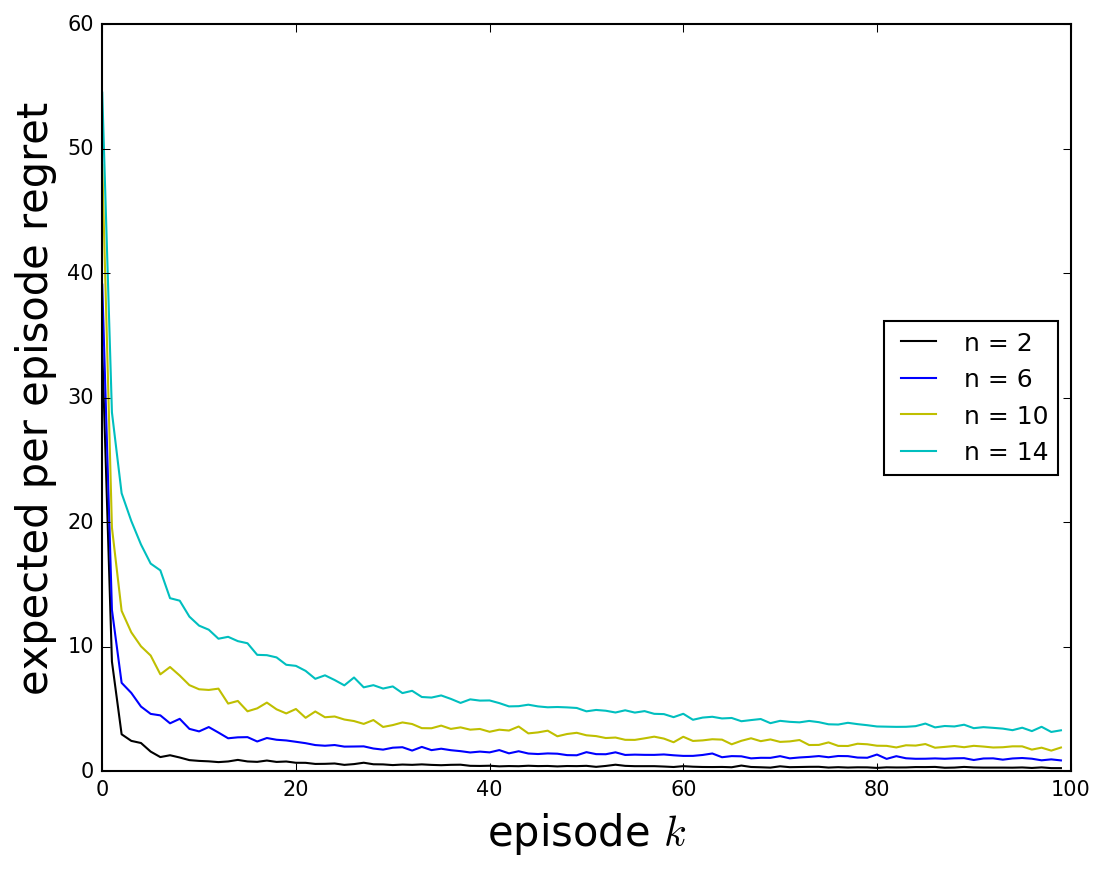}
\caption{Average per-episode regret of  TP for different memory durations $n$.}
\label{fig1b}
\end{figure}

To explore the effect of  memory duration $n$ on the performance of TP, we simulated the same scenario but with different values for $n$. Figure \ref{fig1b} shows the expected  per-episode regret of TP over 100 episodes for $n=2,6,10,14$. As depicted in this figure, TP suffers more regret when $n$ is larger as in this case the prices have more persistent consequences and it takes longer for TP to learn the optimal pricing policy for an episode.

\begin{figure}[t]
\centering
  \includegraphics[width=0.6\linewidth]{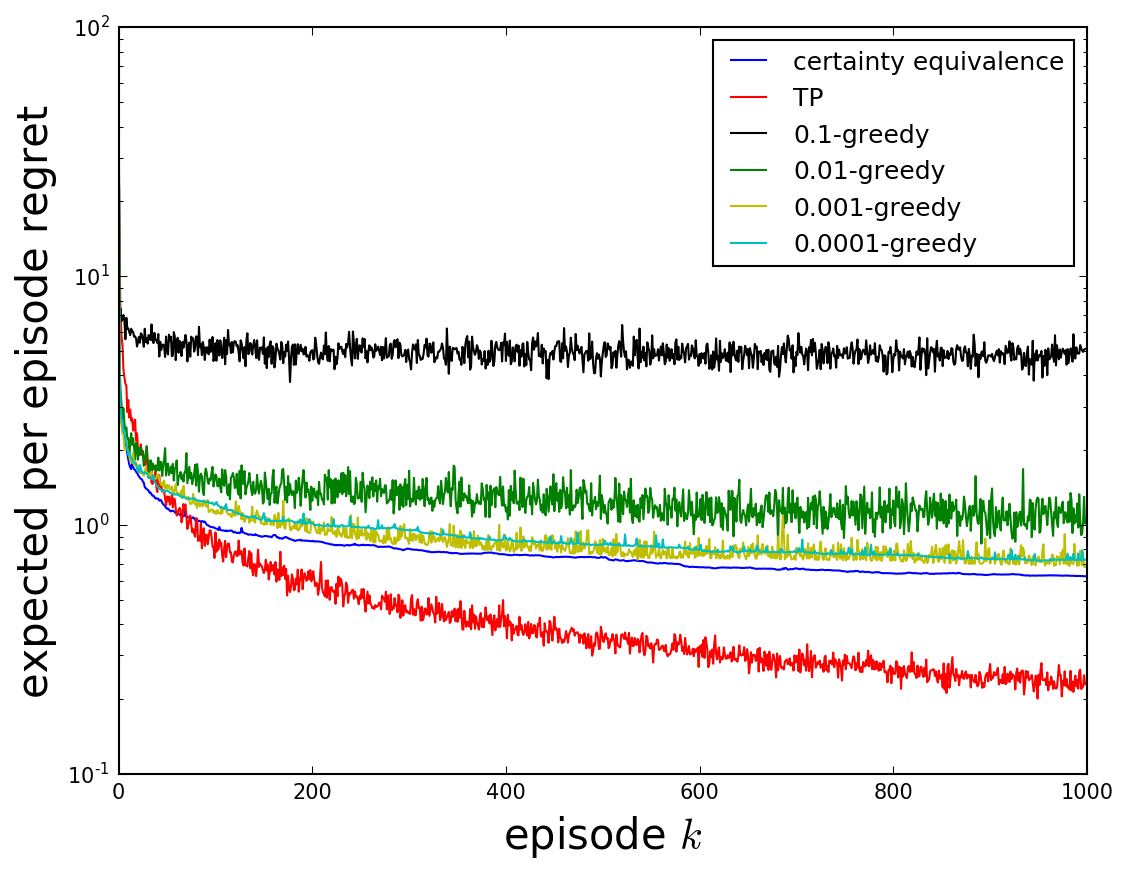}
\caption{Average per-episode regret of  TP compared with certainty equivalence and dithering algorithms.}
\label{fig2}
\end{figure}

An alternative pricing algorithm that is suitable in the described linear environment is  one designed based on certainty equivalence principle. The certainty equivalence  pricing strategy  works similar to TP, except that, instead of sampling from the posterior distribution at the beginning of each episode, it uses the Maximum Likelihood estimate of  $\theta$ to compute  price trajectory within each  episode. Furthermore, to enforce exploration in such an algorithm, dithering techniques, such as $\epsilon$-greedy, can be adopted.  At any period, $\epsilon$-greedy pricing strategy follows  certainty equivalence strategy with probability $1-\epsilon$ and sets a random price with probability $\epsilon$.  To compare the performance of these alternative pricing strategies with that of TP, we  simulate the same scenario  described above. Figure \ref{fig2} shows  average per-episode regret of TP, certainty equivalence strategy and $\epsilon$-greedy strategy for various values of $\epsilon$. The vertical axis in Figure \ref{fig2} is in logarithmic scale to better present the differences. As depicted in this figure, certainty equivalence strategy works reasonably well in the described environment and adding randomness via dithering does not improve its performance. However, TP presents a superior performance as its per-episode regret converges to 0 at a faster rate. Note that in this scenario, the maximum possible price is $1$ and hence the $0.5$ difference between the per-episode regret of TP and certainty equivalence strategy after 1000 episodes presents a significant improvement.

Based on the results of Theorem \ref{mainth}, an upper bound can be established for  cumulative regret of TP when applied in the described linear demand environment.
Before stating the result, we assume that the parameter space is bounded.
\begin{assumption}
\label{assump2}
There exists $\tau>0$ such that $\forall~\tilde\theta\in\Theta:~\|\tilde\theta\|_2\leq\tau$. 
\end{assumption}
The following corollary  follows directly from Theorem \ref{mainth}.
\begin{corollary}
\label{reglinst}
Consider an environment  where expected demand is linearly parameterized as in \eqref{expdemstlin}.  Under Assumptions \ref{revassump} and \ref{assump2}, cumulative regret of TP after $K$ episodes satisfies
\begin{equation}
\label{regstlineq0}
R_K(TP) = O\left(p_{\max}\sigma n^2\sqrt{KH}\log (np_{\max}\tau KH)\right).
\end{equation}
\end{corollary}
\proof{Proof of Corollary \ref{reglinst}.}
Let $\F$ be the class of all possible demand functions as in \eqref{expdemstlin}. It is easy to see that (see, for example, Proposition 2 of \cite{osband2014model})
$$d_K(\F) = O(n^2),~~~d_E(\F,\epsilon) = O(n^2\log(np_{\max}\tau/\epsilon)).$$
Then, the statement  follows from Theorem \ref{mainth}.
\Halmos
\endproof
Corollary \ref{reglinst} shows that per-episode regret of TP in the described linear environment  decreases at a rate of $\log K/\sqrt{K}$ with the number of episodes $K$. Pricing strategies that have been proposed in the literature for memoryless demand models achieve a similar per-episode regret rate in the absence of reference effects  \cite{ferreira2015online}. This indicates that although dynamic pricing with reference effects entails additional challenges, TP  performs efficiently in that context with no additional cost in terms of the regret rate. Moreover, the regret bound established by Corollary \ref{reglinst} is increasing in the history parameter $n$. Clearly, as $n$ increases the prices have more persistent effects and the optimal strategy admits a more complicated structure. Hence, it takes longer for TP to learn the optimal strategy. From another perspective, $n$ dictates the number of unknown parameters of the model and it takes longer to effectively learn within a model with more unknown parameters.  Furthermore, as the horizon $H$ increases, the optimal policy within an episode takes a more complicated form as more sophisticated planning is required for larger horizons. Therefore, as reflected in \eqref{regstlineq0}, it takes longer for TP to effectively learn the optimal policy in larger horizons.

\section{Extensions}
\label{generalSec}
For the sake of exposition of our main ideas, we have so far focused our attention on a simplified setting where indistinguishable products are sold sequentially and our description of TP has been adapted to this scenario. In this section, we discuss how TP can be generalized to incorporate the effect of covariates on the demand and carry out multiproduct pricing possibly with variable and  overlapping life cycles. 

\subsection{The Effect of  Covariates}
\label{covsec}
The setting of Section \ref{setupsec} can be extended to the case where the product being sold at subsequent episodes are  distinct. For example, at an episode the seller may be selling a certain type of coat while in the next episode, he will be selling a certain type of shoe. Although in both episodes the seller deals with the same environment, the coat and the shoe will experience different demands when offered with the same price. More generally, in addition to  prices and  consumer behavior,  demand for a product depends on different characteristics of the product itself. To allow for such dependencies, we assume that at the beginning of episode $k$, the agent has access to a context vector $z_k\in\R^m$ which encodes different characteristics of the product being sold at that episode such as its lifetime, its production cost and whether a similar product currently exists in the market. Furthermore, $z_k$ may contain other covariates which can influence the demand at episode $k$ such as the current inflation rate and the average income of the potential consumers. The context vector may differ from episode to episode while it remains fixed over each episode.

To incorporate the effect of the context on the demand, we extend our formulation in \eqref{expdemstrategic} and assume that the expected demand observed at  period $h$ for product $k$-- the product being sold at episode $k$-- is 
\begin{equation}
\label{demandcov}
d_{k,h} = f_\theta(p_{k,h}, s_{k,h},z_k),
\end{equation}
where similar to Section \ref{setupsec}, $p_{k,h}$ is the price of  product $k$ at  period $h$, $s_{k,h}$ is the $n-$step price history of the product representing the reference effect and $\theta$ is an unknown parameter. In this case, a pricing policy $\pp = (p_1,p_2,\cdots,p_H)\in[0,p_{\max}]^H$ achieves an expected revenue of 
\begin{equation}
\label{valuecov}
V^k_\pp = \sum_{h=1}^H p_h f_\theta(p_h,s_h,z_k)
\end{equation}
in episode $k$, where $s_h$ is the state induced by policy $\pp$ at period $h$. Then, the optimal pricing policy at episode $k$ is $\pp^*_k = \argmax_{\pp}{~V^k_\pp}$. Note that the optimal pricing policy in episode $k$ depends on the context $z_k$. Similar to Section \ref{setupsec} and given the parameter $\theta$, the optimal policy in episode $k$ can be computed by means of a dynamic program. We write 
$$\pp^*_k = \text{DP} (\theta,H,p_{\max},z_k)$$
to indicate that the policy $\pp^*_k$ is the solution of the associated dynamic program at episode $k$ with $H$ periods and given the parameter $\theta$ and context $z_k$. Also, $p_{\max}$ represents the maximum allowable price for the products. 

To better illustrate the extension of TP to this setting, we focus on a linear demand model. Specifically, we assume that the expected demand  at period $h$ in episode $k$ is given by 
\begin{equation}
\label{expdemstlincov}
d_{k,h} = \left\{ \begin{array}{ll}
\alpha + (z_k^\top \beta)p_{k,h}	\qquad  &\text{if }h=1\\
\alpha + (z_k^\top \beta)p_{k,h}+z_k^\top \phi_{h-1} s_{k,h} \qquad
&\text{if }2\leq h\leq n\\
\alpha + (z_k^\top \beta)p_{k,h}+z_k^\top \phi_{n} s_{k,h}	 \qquad
&\text{if }n+1\leq h
\end{array}
\right .
\end{equation}
where $\alpha\in\R$, $\beta\in\R^m$ and $\forall~1\leq i\leq n:~\phi_i\in\R^{m\times i}$ are  unknown  parameters of the demand function. There are a total of $1+m+m{n(n+1)}/{2}$ unknown parameters in this model which can be encoded in terms of a vector 
$\theta = \begin{bmatrix}
\alpha,\beta^\top,\bar \phi_1^\top,\cdots,\bar \phi_n^\top\end{bmatrix}^\top,$
where $\bar \phi_i$ is an $mi$ dimensional vector generated by stacking the columns of $\phi_i$ into a single column. 

TP can be adopted in the same way as in Section \ref{specialSec} to generate pricing policies for this problem. Specifically, starting with a $N(\mu,\Sigma)$ prior distribution  on $\theta$, TP draws a sample from the posterior distribution at the start of each episode and uses dynamic programming to compute a policy which is then executed throughout the episode. Thanks to conjugacy properties of normal distributions, the posterior distribution of $\theta$ after any number of episodes remains normal. To specify the update rules for posterior means and covariances, let us define some auxiliary variables as in Section \ref{specialSec}. Given the observations $\left(z_k,p_{k,1},s_{k,1},y_{k,1},\cdots,p_{k,H},s_{k,H},y_{k,H}\right)$ gathered at episode $k$, define $$
w_{k,h} = \log(y_{k,h}) + \frac{\sigma^2}{2},$$
and for any $1\leq h\leq H$, let
$$x_{k,h} = \left\{ \begin{array}{ll} 
 \big[1,p_{k,h} z_k^\top, \overbrace{0\cdots,0}^{\frac{mn(n+1)}{2}}\big] \qquad & \text{if  } h=1\\
 \big[1,p_{k,h} z_k^\top,\overbrace{0\cdots,0}^{\frac{m(h-1)(h-2)}{2}}, (s_{k,h}\otimes z_k)^\top, \overbrace{0\cdots,0}^{\frac{m(n(n+1)-h(h-1))}{2}}\big] \qquad & \text{if  } 2\leq h\leq n\\
\big[1,p_{k,h} z_k^\top,\overbrace{0\cdots,0}^{\frac{mn(n-1)}{2}}, (s_{k,h}\otimes z_k)^\top\big] \qquad & \text{if  } h\geq n+1 ,
\end{array}\right .$$
where $\otimes$ denotes the Kronecker product.
 Then, at the end of episode $k$, the mean and covariance matrix of the posterior distribution are updated according to 
\begin{equation}
\label{update2cov}
\mu\leftarrow \left(\Sigma^{-1} + \frac{1}{\sigma^2}\sum_{h=1}^H x_{k,h}^\top x_{k,h}\right)^{-1}\left(\Sigma^{-1}\mu + \frac{1}{\sigma^2}\sum_{h=1}^H w_{k,h} x_{k,h}^\top\right),~~~~\Sigma\leftarrow \left(\Sigma^{-1} + \frac{1}{\sigma^2}\sum_{h=1}^H x_{k,h}^\top x_{k,h}\right)^{-1}.
\end{equation}

\begin{algorithm}[t]
   \caption{\texttt{DP-lin-cov}}
   \label{alg5}
\begin{algorithmic}
   \STATE {\bfseries Input:} $\theta,H,p_{\max},z$
   \STATE {\bfseries Output:} $\pp$
   \STATE Extract $\alpha,\beta,\phi_1,\cdots,\phi_n$ from $\theta$
   \STATE{\bfseries Initialize:} $M = \mathbf{0}_{H\times H}$
   \STATE Set $M[1,1] = z^\top\beta$
   \FOR{$h=2,3,\cdots,H$}   
   \STATE Set $M[h,h]=z^\top\beta$
   \STATE Set $M[\max(h-n,1):h-1,h] = \frac{1}{2}z^\top \phi_{\max(h-1,n)}$
   \STATE Set $M[h,\max(h-n,1):h-1] = \frac{1}{2}\phi_{\max(h-1,n)}^\top z$
   \ENDFOR
   \STATE Return $\pp=\argmax_{x\in[0,p_{\max}]^H}{~
x^\top Mx + \alpha\mathbf{1}^\top x}$
\end{algorithmic}
\end{algorithm} 

Note that the state in any time period is determined by the state in the previous time period and the selected price. Thanks to such deterministic evolution of the states and linearity of the demand function, the dynamic program step in TP reduces to a quadratic optimization problem. To see this, note that from \eqref{valuecov} and \eqref{expdemstlincov}, the expected revenue of a price vector $\pp = [p_1,p_2,\cdots,p_H]^\top\in[0,p_{\max}]^H$ in the $k$'th episode can be expressed as 
\begin{equation}
\label{multivalueexpand}
V^k_\pp = \sum_{h=1}^H \alpha p_h + \sum_{h=1}^H (z_k^\top\beta)p_h^2 + \sum_{h=2}^n p_h (z_k^\top\phi_{h-1}s_h) + \sum_{h=n+1}^H p_h (z_k^\top\phi_{n}s_h).
\end{equation}
Now, given the set of parameters $\theta$, define the $H\times H$ matrix $M_\theta$ such that
\begin{enumerate}
\item $M_\theta[h,h] = z_k^\top\beta$ for $1\leq h\leq H$,
\item $M_\theta[\max(h-n,1):h-1,h] = \frac{1}{2}z_k^\top\phi_{\max(h-1,n)}$ for all $2\leq h\leq H$,
   \item $M_\theta[h,\max(h-n,1):h-1] = \frac{1}{2}\phi_{\max(h-1,n)}^\top z_k$ for all $2\leq h\leq H$,
\item all other entries of $M$ are equal to 0.
\end{enumerate}
Then, \eqref{multivalueexpand} can be expressed as
\begin{equation}
\label{multivalueexpand2}
V_\pp^k = \pp^\top M_\theta \pp + \alpha \1^\top \pp.
\end{equation}
Therefore, given the sampled parameter $\hat \theta_k$ at the beginning of episode $k$, the policy to be applied in episode $k$ is given by
\begin{equation}
\label{quadopt2}
{\pp}_k = \argmax_{\pp\in[0,p_{\max}]^H}{~\pp^\top M_{\hat\theta_k} \pp + \hat\alpha_k \1^\top \pp}.
\end{equation}

Similar to Section \ref{specialSec}, the matrix $M_{\hat\theta_k}$ is not guaranteed to be negative semi-definite in which case the optimization problem in \eqref{quadopt2} is not convex. However, with appropriate choice of the prior mean (for example when $z_k^\top\hat\beta_k$ is negatively large with high probability) $M_{\hat\theta_k}$ would be negative semi-definite with high probability. When implementing TP, posterior sampling at the beginning of episode $k$ can be repeated until the sampled $\hat\theta_k$ results in a negative semi-definite $M_{\hat\theta_k}$. Algorithm \ref{alg5} describes successive steps of the above solution and Algorithm \ref{alg6} presents the generalization of $TP$ to incorporate the effect of covariates.

\begin{algorithm}[t!]
   \caption{\texttt{TP-lin-cov}}
   \label{alg6}
\begin{algorithmic}
   \STATE {\bfseries Input:} $H,p_{\max},\mu_0,\Sigma_0,\sigma^2$
   \STATE{\bfseries Initialize:} $\mu=\mu_0,\Sigma=\Sigma_0$
   \FOR{$k=1,2,\cdots$}   
   \STATE Observe context variable $z_k$
   \STATE Sample $\hat\theta_k\sim N(\mu,\Sigma)$
   \STATE Compute $\hat\pp_k=(p_{k,1},p_{k,2},\cdots,p_{k,H})=\texttt{DP-lin-cov}(\hat\theta_k,H,p_{\max},z_k)$
   \FOR{$h=1,2,\cdots,H$}
   \STATE Set price $p_{k,h}$
   \STATE Observe random demand $y_{k,h}$
   \ENDFOR
   \STATE Update $\mu$ and $\Sigma$ according to \eqref{update2cov}
   \ENDFOR
\end{algorithmic}
\end{algorithm} 

We can also generalize the result of Theorem \ref{mainth} and establish a regret bound for TP in such a scenario. Let $\Z$ denote the set of all possible context vectors. We make the following assumption.
\begin{assumption}
\label{assump3}
There exists $\lambda>0$ such that $\forall~z\in\Z:~\|z\|_2\leq \lambda$. 
\end{assumption}
The following theorem provides a regret bound for TP in the above linear  environment. 
\begin{theorem}
\label{reglinstcov}
Consider an environment  where the expected demand is given as in \eqref{expdemstlincov}. Under Assumptions \ref{revassump}, \ref{assump2} and \ref{assump3}, the regret of TP after $K$ episodes  would be 
\begin{equation}
\label{regstlineq}
R_K(TP) = O\left(p_{\max}H\sigma mn^2\sqrt{KH}\log(np_{\max}\lambda\tau KH)\right).
\end{equation}
\end{theorem}
Theorem \ref{reglinstcov} has been proved in Section \ref{analysissection}. This Theorem  yields a similar regret bound for TP in the presence of the covariates as in the case of no covariates. The only  difference is the dependence of \eqref{regstlineq} on the number of covariates $m$. Note that $m$  scales the number of unknown parameters of the model and  appears lineraly in the regret bound.

\subsection{Multiproduct Pricing}
\label{multisection}
So far, we have been considering a single product dynamic pricing problem where at each episode, the seller prices and sells a single product. In many practical situations, however, multiple products are being sold by the seller and he needs to simultaneously price all of them. Potentially, these products can be related to each other in a way that the demand for one of them depends on the price of all the products. 
Specifically, suppose that a set of $q$ products are being sold at episode $k$. At period $h$ in this episode, the agent sets a price vector $P_{k,h}\in[0,p_{\max}]^q$  such that its $j$'th component, denoted as $P_{k,h}[j]$, is the price of  product $j$. While we can extend the model to incorporate the effect of covariates on the demand as well, we neglect such effects here to ease the exposition. Let 
$$S_{k,h} =\begin{bmatrix}
 P_{k,{\max(1,h-n)}}\\
 \vdots\\
 P_{k,h-2}\\
 P_{k,h-1}\\
 \end{bmatrix}$$
be the $q\min(n,h-1)$ dimensional state vector at period $h$ in episode $k$ and let $D_{k,h}$ be the demand vector at this period such that $D_{k,h}[j]$ is the expected demand for product $j$.  
Focusing on a linear demand function, the expected demand  can be modeled as
\begin{equation}
\label{expdemstlincovmulti}
D_{k,h} = \left\{ \begin{array}{ll}
\alpha +  \beta P_{k,h}	\qquad  &\text{if }h=1\\
\alpha +  \beta P_{k,h}+ \phi_{h-1} S_{k,h} \qquad
&\text{if }2\leq h\leq n\\
\alpha + \beta P_{k,h}+ \phi_{n} S_{k,h}	 \qquad
&\text{if }n+1\leq h,
\end{array}
\right .
\end{equation}
 where $\alpha\in\R^q$, $\beta\in\R^{q\times q}$ and $\forall ~1\leq i\leq n: \phi_i\in\R^{q\times q\min(n,i)}$ are the unknown  parameters of the demand function. There are a total of $q+q^2+q^2{
 n(n+1)}/{2}$ unknown parameters which can be encoded in a vector $\theta = \left[\alpha^\top,\bar\beta^\top,\bar\phi_1^\top,\cdots,\bar\phi_n^\top\right]^\top$ where $\bar\beta$ is a vector generated by stacking the columns of $\beta$ on top of each other and $\bar\phi_i$ is generated in the same way for $1\leq i\leq n$. 
 
 The agent observes a random demand vector  $Y_{k,h}$ at period $h$ in episode $k$ such that $Y_{k,h}[j]$, the demand observed for product $j$, is a log-normal random variable with parameters $D_{k,h}[j] - {\sigma^2}/{2}$ and $\sigma^2$. Note that we have $\E[Y_{k,h}] = D_{k,h}$. The expected revenue achieved at this period is $P_{k,h}^\top D_{k,h}$. A pricing policy in this case is a sequence of price vectors $\PP=(P_1,\cdots,P_H)$ such that $\forall 1\leq h\leq H:~P_h\in [0,p_{\max}]^q$. The optimal pricing policy is the one that maximizes the expected revenue over the episode:
 \begin{equation}
 \label{optvaluemulti}
 \PP^* = \argmax_{\PP} \sum_{h=1}^H P_h^\top D_h.
 \end{equation}
 Similar to the single product setting, this optimization problem can be solved via a dynamic program. We overload our notation and write 
 \begin{equation}
 \label{dpmulti0}
 \PP^* = \mbox{DP}(\theta,H,p_{\max},q)
 \end{equation}
 to denote that $\PP^*$ is the solution of the optimization problem in \eqref{optvaluemulti} when the demand function is governed by the parameter $\theta$. 
 \begin{algorithm}[t]
   \caption{\texttt{DP-lin-mult}}
   \label{alg7}
\begin{algorithmic}
   \STATE {\bfseries Input:} $\theta,H,p_{\max},q$
   \STATE {\bfseries Output:} $\PP$
   \STATE Extract $\alpha,\beta,\phi_1,\cdots,\phi_n$ from $\theta$
   \STATE{\bfseries Initialize:} $M = \mathbf{0}_{qH\times qH}$
   \STATE Set $M[1:q,1:q] = \beta$
   \FOR{$h=2,3,\cdots,H$}   
   \STATE Set $M[(h-1)q+1:hq,(h-1)q+1:hq]=\beta$
   \STATE Set $M[\max((h-n-1)q,0)+1:(h-1)q,h] = \frac{1}{2}\phi_{\max(h-1,n)}^\top$
   \STATE Set $M[h,\max((h-n-1)q,0)+1:(h-1)q] = \frac{1}{2}\phi_{\max(h-1,n)} $
   \ENDFOR
   \STATE Build matrix $A$ by stacking $\alpha$ over itself $H$ times
   \STATE Find $x^*=\argmax_{x\in[0,p_{\max}]^{qH}}{~
x^\top Mx + A^\top x}$
\FOR{$h=1,2,\cdots,H$}
\STATE Set $P_h = x^*[(h-1)q+1:hq]$
\ENDFOR
\STATE Return $\PP = (P_1,\cdots,P_H)$
\end{algorithmic}
\end{algorithm} 

TP can be adapted to learn the optimal policy in this multiproduct pricing problem. 
Similar to single product scenario, TP  starts with a $N(\mu,\Sigma)$ prior distribution on $\theta$. Similar to Section \ref{specialSec} and thanks to conjugacy properties of normal distributions, the posterior distribution of $\theta$ after any number of episodes remains normal. To specify the update rules for posterior means and covariances, we define some auxiliary variables. Upon observing $(P_{k,1},Y_{k,1},P_{k,2},S_{k,2},Y_{k,2},\cdots,P_{k,H},S_{k,H},Y_{k,H})$ at episode $k$, let $W_{k,h}\in\R^H$ be 
$$\forall~1\leq j\leq q:~W_{k,h}[j] = \log(Y_{k,h}[j]) + \frac{\sigma^2}{2},$$
and for any $1\leq h\leq H$, define $X_{k,h}$ as
$$X_{k,h} = \left\{ \begin{array}{ll} 
 \big[I_q,P_{k,h}^\top\otimes I_q , \mathbf{0}_{q\times q^2\frac{n(n+1)}{2}} \big] \qquad & \text{if  } h=1\\
 \big[I_q,P_{k,h}^\top\otimes I_q , \mathbf{0}_{q\times q^2\frac{(h-1)(h-2)}{2}}, S_{k,h}^\top\otimes I_q,\mathbf{0}_{q\times q^2\frac{n(n+1)-h(h-1)}{2}} \big] \qquad & \text{if  } 2\leq h\leq n\\
\big[I_q,P_{k,h}^\top\otimes I_q , \mathbf{0}_{q\times q^2\frac{n(n-1)}{2}}, S_{k,h}^\top\otimes I_q\big] \qquad & \text{if  } h\geq n+1 .
\end{array}\right .$$
Some linear algebra leads to the following update rules for the mean and covariance matrix at the end of episode $k$:
\begin{equation}
\label{muSigmaupdate3}
\mu\leftarrow\left(\Sigma^{-1} + \frac{1}{\sigma^2}\sum_{h=1}^H X_{k,h}^\top X_{k,h}\right)^{-1}\left(\Sigma^{-1}\mu + \frac{1}{\sigma^2} \sum_{h=1}^H X_{k,h}^\top W_{k,h}\right),~~~~~\Sigma\leftarrow \left(\Sigma^{-1} + \frac{1}{\sigma^2}\sum_{h=1}^H X_{k,h}^\top X_{k,h}\right)^{-1}.
\end{equation} 

 At the beginning of each $k$'th episode, TP draws a sample $\hat\theta_k$ from the posterior distribution and  treating it as the truth, computes the policy
 \begin{equation}
 \label{dpmulti}
 \PP_k = \mbox{DP}(\hat\theta_k,H,p_{\max},q)
 \end{equation}
 and applies it throughout the episode.
 
Note that similar to Section \ref{specialSec}, the state vectors in the described multiproduct pricing model evolve deterministically. Due to this fact and  linearity of the demand function in \ref{expdemstlincovmulti}, the dynamic program in \ref{dpmulti0} reduces to a quadratic optimization problem as described in Algorithm \ref{alg7}. The proof of this reduction is similar to that presented in Section \ref{specialSec} for the single product scenario and is omitted here. Algorithm \ref{alg8} describes the generalization of TP to the above multiproduct pricing setting.

\begin{algorithm}[t!]
   \caption{\texttt{TP-lin-mult}}
   \label{alg8}
\begin{algorithmic}
   \STATE {\bfseries Input:} $H,q,p_{\max},\mu_0,\Sigma_0,\sigma^2$
   \STATE{\bfseries Initialize:} $\mu=\mu_0,\Sigma=\Sigma_0$
   \FOR{$k=1,2,\cdots$}   
   \STATE Sample $\hat\theta_k\sim N(\mu,\Sigma)$
   \STATE Compute $\hat\PP_k=(P_{k,1},P_{k,2},\cdots,P_{k,H})=\texttt{DP-lin-mult}( \hat\theta_k,H,p_{\max},q)$
   \FOR{$h=1,2,\cdots,H$}
   \STATE Set price vector $P_{k,h}$
   \STATE Observe random demand $Y_{k,h}$
   \ENDFOR
   \STATE Update $\mu$ and $\Sigma$ according to \eqref{muSigmaupdate3}
   \ENDFOR
\end{algorithmic}
\end{algorithm} 

Thanks to our general analysis in Section \ref{analysissection}, we can also provide a regret bound for TP in the described multiproduct pricing problem. As a generalization of Assumption \ref{revassump}, we make the following assumption on demand vectors.
\begin{assumption}
\label{assump4}
There exists  constant $d_{\max}>0$ such that all the demand vectors $D$  satisfy $\|D\|_\infty\leq d_{\max}$. 
\end{assumption}
The following theorem provides a regret bound  for TP.
\begin{theorem}
\label{multtheorem}
Consider a  multiproduct pricing problem where  the expected demand is given by \eqref{expdemstlincovmulti}. Under Assumptions \ref{assump2} and \ref{assump4}, the regret of TP after $K$ episodes would be
\begin{equation}
\label{multtheoremeq1}
R_K(TP) = O\left(p_{\max} \sigma q^3 n^2 \sqrt{ KH}\log(\tau KH)\right).
\end{equation}
\end{theorem}
Theorem \ref{multtheorem} has been proved in Section \ref{analysissection}. The only difference between the above regret bound and the one established in Corollary \ref{reglinst} for the single product scenario is the appearance of the number of products $q$ in the regret bound. As the number  of products increases, TP requires more time to effectively learn the influence of a product's price on  demand for other products. 

\subsection{Asynchronous Product Pricing}
As another interesting scenario, we consider dynamic pricing of multiple products with variable and overlapping life cycles. Particularly, we allow the seller to start selling  other products while he is still busy with selling others. Let $t=1,2,\cdots$ be the time index denoting the time since the seller's marketing campaign has started. Suppose that product $k$ is launched at time $t_k$ and needs to be sold  in $H_k$ consecutive time periods (i.e., until time $t_k+H_k-1$). We refer to the time interval $[t_k,t_k+H_k-1]$ as episode $k$ which consists of $H_k$ periods and might  overlap with other episodes. Note that period $h$ at episode $k$ corresponds to time index  $t_{k,h}=(t_k+h-1)$. We assume that at the beginning of episode $k$, the seller has access to a context vector $z_k\in\R^m$ which encodes the specific characteristics of product $k$ (the product being sold at episode $k$) and other covariates. Similar to  Subsection \ref{covsec}, we assume that the expected demand at period $h$ in episode $k$ is given by
\begin{equation}
\label{expdemstlincovassync}
d_{k,h} = \left\{ \begin{array}{ll}
\alpha + (z_k^\top \beta)p_{k,h}\qquad  &\text{if }h=1\\
\alpha + (z_k^\top \beta)p_{k,h}+z_k^\top \phi_{h-1} s_{k,h} \qquad
&\text{if }2\leq h\leq n\\
\alpha + (z_k^\top \beta)p_{k,h}+z_k^\top \phi_{n} s_{k,h}	 \qquad
&\text{if }n+1\leq h\leq H_k,
\end{array}
\right .
\end{equation}
where $p_{k,h}$ is the price of product $k$ at period $h$ in episode $k$ and $s_{k,h} = [p_{k,\max(1,h-n)},\cdots,p_{k,h-1}]^\top$ is the price history of product $k$ at this time. Similar to Subsection \ref{covsec}, the demand function has $1+m+m{n(n+1)}/{2}$ unknown parameters which can be encoded in a vector $\theta = \begin{bmatrix}
\alpha,\beta^\top,\bar \phi_1^\top,\cdots,\bar \phi_n^\top\end{bmatrix}^\top$. For the sake of consistency, we  assume that $n\leq H_k$ for all $k\in\N$. 
At period $h$ in episode $k$, a demand $y_{k,h}$ is observed for product $k$ which is a log-normal random variable with parameters $d_{k,h}-{\sigma^2}/{2}$ and $\sigma^2$. This indicates that the products are disjoint; that is, the price of a product does not affect the demand for other products being sold at the same time.

TP can be easily adapted to such an asynchronous pricing problem. In this case, TP starts with a $N(\mu,\Sigma)$ prior distribution on $\theta$. Again and due to the conjugacy properties of normal distributions, the posterior distribution of $\theta$ after each time period and for any number of products being sold at the same time remains normal. In this case, TP updates the posterior parameters based on the gathered   observations after each period in order to maintain the most up to date posterior distribution at any time. Once the new product $k$ with context $z_k$ is launched and episode $k$ of length $H_k$ starts, TP draws a sample $\hat\theta_k$ from the prevailing posterior distribution, computes the policy 
$$\pp_k = \mbox{DP}(\hat\theta_k,H_k,p_{\max},z_k),$$
and applies it throughout the episode. 

The update rules for posterior means and covarainces after each period can be derived similar to Section \ref{covsec}. The only difference is that at any given period, there might be observations associated to multiple products in which case the observations are aggregated  and used to update the posterior parameters. Algorithm \ref{alg10} presents a detailed procedure for updating the posterior parameters after each period. Successive steps of TP adapted to the above asynchronous pricing scenario are described in Algorithm \ref{alg9}.

\begin{algorithm}[t!]
   \caption{\texttt{update}}
   \label{alg10}
\begin{algorithmic}
   \STATE {\bfseries Input:} $\mu,\Sigma,z,\pp,h,y,\sigma^2,n,m$
   \STATE Consider $\pp$ as $\pp = [p_1,p_2,\cdots,p_H]$ 
   \IF{$h=1$}
   \STATE Let 
   $x = [1,p_h z^\top,\mathbf{0}_{1\times mn(n+1)/2}]$
   \ELSIF{$2\leq h\leq n$}
\STATE Let $s =[p_1,p_2,\cdots,p_{h-1}]$    
    \STATE Let 
   $x = [1,p_h z^\top,\mathbf{0}_{1\times m(h-1)(h-2)/2},(s\otimes z)^\top,\mathbf{0}_{1\times m(n(n+1)-h(h-1))/2}]$
   \ELSE
  \STATE Let $s =[p_{h-n},\cdots,p_{h-2},p_{h-1}]$    
    \STATE Let 
   $x = [1,p_h z^\top,\mathbf{0}_{1\times mn(n-1)/2},(s\otimes z)^\top]$
   \ENDIF
   
   \STATE Let $w = \log(y)+\frac{\sigma^2}{2}$
   \STATE Define 
   $$\mu_{\text{new}}= \left(\Sigma^{-1} + \frac{1}{\sigma^2} x^\top x\right)^{-1}\left(\Sigma^{-1}\mu + \frac{1}{\sigma^2} w x^\top\right),~~~~\Sigma_{\text{new}}= \left(\Sigma^{-1} + \frac{1}{\sigma^2} x^\top x \right)^{-1}$$
   
   \STATE Return $\mu_{\text{new}}$ and $\Sigma_{\text{new}}$
\end{algorithmic}
\end{algorithm}

\begin{algorithm}[t!]
   \caption{\texttt{TP-lin-asynch}}
   \label{alg9}
\begin{algorithmic}
   \STATE {\bfseries Input:} $p_{\max},\mu_0,\Sigma_0,\sigma^2,n,m$
   \STATE{\bfseries Initialize:} $\mu=\mu_0,\Sigma=\Sigma_0, A =\emptyset, e=1$
   \FOR{$t=1,2,3,\cdots$}
   \WHILE{a new product  is launched}
   \STATE Observe the episode length $H_e$ and covariates $z_e\in\R^m$
   \STATE Set the starting position of episode $e$ as $t_e=t$
   \STATE Sample $\hat\theta_e\sim\N(\mu,\Sigma)$
   \STATE Compute $\pp_k = \texttt{DP-lin-cov}(\hat\theta_e,H_e,p_{\max},z_k)$
   \STATE Add $e$ to $A$ and $e\leftarrow e+1$
   \ENDWHILE
   \FOR{$k\in A$}
   \STATE Let $h=t-t_k+1$ be the period index in episode $k$
   \STATE Set price $p_{k,h}$ for product $k$
   \STATE Observe random demand $y_{k,h}$ 
   \STATE Update the parameters $\mu,\Sigma\leftarrow \texttt{update} (\mu,\Sigma,z_k, \pp_k,h,y_{k,h},\sigma^2,n,m)$
   \ENDFOR
   \ENDFOR
\end{algorithmic}
\end{algorithm}

\section{Analysis of TP}
\label{analysissection}
Our proposed TP algorithm is an adaptation of  PSRL algorithm proposed in \cite{ian1}, and \cite{osband2014model} provides a regret bound on its performance when employed in a general reinforcement learning problem. However, this general regret bound does not take into account the special structure of the dynamic pricing problem. Particularly,  the regret bound derived in \cite{osband2014model} depends on a Lipschitz constant ($K^*$ in equation (4) of \cite{osband2014model}) which is not easy to quantify in the reinforcement learning problem of our interest. Instead, we take an approach similar to \cite{dan1} and \cite{osband2014model}, and exploiting the special structure of the dynamic pricing problem, derive a regret bound on the performance of TP.

To maintain generality, we analyze the performance of TP in an environment with reference effects and in the presence of covariates where multiple products are being sold at each episode. In te remaining of this section, We first describe this general scenario and then state the main result of the paper which is a regret bound for TP in this scenario followed by its proof. Finally and given this general result, we prove the regret bounds established in Theorems \ref{mainth}, \ref{reglinstcov} and \ref{multtheorem}.  

\subsection{A General Scenario}
Consider a  seller who markets for and sells $q$ products at each episode consisting of $H$ periods. At the beginning of episode $k$, a context $z_k\in\R^m$ is available to the seller which encodes characterisitcs of the products being sold and other covariates related to episode $k$.
At period $h$ in episode $k$, the seller selects a price vector $P_{k,h} \in [0,p_{\max}]^q$ such that $P_{k,h}[j]$ denotes the price set for product $i$ at this period, $1\leq i\leq q$. The demand  for a product at any time may depend on the prevailing and previous prices of its own as well those of other products. To represent such reference effects, we let 
$$S_{k,h} =\begin{bmatrix}
 P_{k,\max(1,h-n)}\\
 \vdots\\
 P_{k,h-2}\\
 P_{k,h-1}\\
 \end{bmatrix}$$
be the state at period $h$ in episode $k$ which is built by concatenating  (at most) $n$ previous price vectors in the same episode. The expected demand vector at period $h$ in episode $k$ can be considered as
\begin{equation}
\label{expdem10}
D_{k,h} = f_\theta(P_{k,h},S_{k,h},z_k),
\end{equation}
for some parameteric function $f_\theta$ such that $D_{k,h}[j]$ represents the expected demand for product $j$ at this period. 
Let $Y_{k,h}$ be the random demand vector experienced by the seller at period $h$ in episode $k$ and assume that $Y_{k,h}[j]$-- random demand observed for product $j$ at this period-- is  log-normally distributed with parameters $D_{k,h}[j] -{\sigma^2}/{2}$ and $\sigma^2$. 
 To maintain a compact notation, when appropriate,  we use $f^k_\theta(P_{k,h},S_{k,h})$ as a shorthand for $f_\theta(P_{k,h},S_{k,h},z_k)$. 
 
 A pricing policy for an episode is a sequence of $H$ price vectors such as $\PP = (P_1,\cdots,P_H)$ such that $P_h\in[0,p_{\max}]^q$. The value of  policy $\PP$ at episode $k$ is 
$$ V^k_\PP =\sum_{h=1}^H P_h^\top D_{h} = \sum_{h=1}^H P_h^\top f^k_{\theta}(P_h,S_h),$$ 
where $S_h$ and $D_h$ 
are the state and expected demand vector induced by policy $\PP$ at period $h$, respectively. Given $z_k$ and $\theta$, the optimal pricing policy for episode $k$ is $\PP^*_k =\argmax_{\PP}{~V^k_\PP},$. 

\subsection{Main Result}
Note that the above formulation of the dynamic pricing problem is general enough to capture all the special scenarios discussed in Sections  \ref{setupsec}, \ref{covsec} and \ref{multisection}. The implementation of TP in each of these scenarios can be thought of as a special case of the following description of TP. 
TP starts with a prior distribution on $\theta$ and, given the observed demands, updates it to a posterior distribution at the end of each episode. At the start of episode $k$, TP  samples $\hat\theta_k$ from the prevailing posterior distribution, computes a policy $\PP_k$ that maximizes 
$$ \hat V^k_\PP = \sum_{h=1}^H P_h^\top f^k_{\hat\theta_k}(P_h,S_h)$$
and applies it throughout the episode. The expected regret of TP after $K$ episodes is 
$$R_K(TP) = \E\left[\sum_{k=1}^K  \left(V^k_{\PP^*_k} - V^k_{\PP_k}\right)\right],$$
where the expectation is taken over the context vectors, the inherent randomness in TP and the randomness in $\theta$ itself. 

Let $\Theta$ be the collection of all possible parameters of the demand function (i.e., the support of the prior distribution) and let
$$\F = \{f_{\theta}:[0,p_{\max}]^q\times \S\times \Z\to \R^q|\theta\in\Theta\}$$
denote the class of demand functions spanned by $\theta\in\Theta$, where $\S = \emptyset\cup_{i=1}^n [0,p_{\max}]^{qi}$ is the state space and $\Z$ denotes the set of all possible context vectors. 
The following theorem, which is the main technical result of this paper, provides a bound on the expected regret of TP in the abovementioned scenario.
\begin{theorem}
\label{mainth0}
Consider an environment with reference effects where $q$ products are being sold at each episode and the expected demand vector at any time is given by \eqref{expdem10}. Assume that the expeted demand vectors always satisfy $\|D\|_\infty\leq d_{\max}$. Then,  the regret of TP after $K$ episodes would be
\begin{equation}
\label{mainth0eq1}
R_K(TP)\leq qp_{\max}\left(1 + Hd_{\max} d_E(\F,(KH)^{-2}) + 4\sqrt{\beta_Kd_E(\F,(KH)^{-2})KH}\right) + \frac{4qp_{\max}d_{\max}}{KH},
\end{equation}
where 
\begin{equation}
\label{mainth0eq2}
\beta_K = 8\sigma^2\log((KH)^2 N(\F,(KH)^{-2}))+\frac{2}{KH} \left(8d_{\max} + \sqrt{8\sigma^2\log 4}\right).
\end{equation}
In an asymptotic notation, the expected regret of TP satisfies
\begin{equation}
\label{mainth0eq3}
R_K(TP) = O\left(qp_{\max}\sigma\sqrt{d_K(\F)d_E(\F,(KH)^{-2})KH\log(KH)}\right).
\end{equation}
\end{theorem}
We provide a proof for Theorem \ref{mainth0} in three steps. 
We start by deriving an upper bound on the regret of TP at each episode. Then, we build a series of delicate confidence sets for the unknown parameter $\theta$ and using the eluder dimension of $\F$  bound the cumulative {\em width} of these confidence sets. Finally, we combine the two previous steps and prove Theorem \ref{mainth0}.

\subsubsection{Bounding the Regret at One Episode}
Let  $\Delta_k = \E\left[V^k_{\PP^*_k} - V^k_{\hat\PP_k} \right]$ denote the expected regret of TP at episode $k$. Also for any $k\in\N$, let $\HH_k=\{(P_{j,1},S_{j,1},Y_{j,1},\cdots,P_{j,H},S_{j,H},Y_{j,H})\}_{j=1}^{k-1}$ be the observations made up to the beginning of episode $k$. 
The following lemma gives an alternative expression for $\Delta_k$ which simplifies the rest of our analysis.
\begin{lemma}
\label{lem1}
for any $k\in\N$, we have
\begin{equation}
\label{Delta2}
\Delta_k=\E\left[ \hat V^k_{\PP_k} - V^k_{\PP_k} \right].
\end{equation}
\end{lemma}
\proof{Proof of Lemma \ref{lem1}.}
By the definition of $\Delta_k$ and by tower property, we can write for any $k$:
\begin{align*}
\Delta_k & = \E\left[\E\left[V^k_{\pp^*_k} - V^k_{\pp_k}\big|\HH_k\right]\right]\\
& =  \E\left[\E\left[V^k_{\pp^*_k} - \hat V^k_{\pp_k}\big|\HH_k\right] + \E\left[\hat V^k_{\pp_k} - \hat V^k_{\pp_k}\big|\HH_k\right]\right].
\end{align*}
On the other hand, since $\theta$ and $\hat\theta_k$ are identically distributed given the history  $\HH_k$, then we have
$$\E\left[V^k_{\pp^*_k} - \hat V^k_{\pp_k}\big|\HH_k\right]  = 0.$$
combining the two above equations proves the statement.
\Halmos
\endproof

Now, let $P_{k,h}$ be the price vector selected by TP at period $h$ in episode $k$ and let $S_{k,h}$ denote the state observed at that period. Where appropriate we may let $U_{k,h}$ denote the pair $(P_{k,h},S_{k,h})$.  The following lemma is a direct consequence of Lemma \ref{lem1}.

\begin{lemma}
\label{lem15}
For any $k\in\N$, we have
\begin{equation}
\label{Del2}
\Delta_k \leq qp_{\max}\E\left[\sum_{h=1}^H \left\| f^k_{\hat \theta_k}\left(U_{k,h}\right) - f^k_{\theta}\left(U_{k,h}\right)\right\|_2\right],
\end{equation}
\end{lemma}
\proof{Proof of Lemma \ref{lem15}.}
From the definition of $V$ and $\hat V$ and using Lemma \ref{lem1}, we have
\begin{align}
\Delta_k &=  \E\left[\sum_{h=1}^H\left[P_{k,h}^\top f^k_{\hat \theta_k}\left(U_{k,h}\right) -  P_{k,h}^\top f^k_{\theta}\left(U_{k,h}\right)\right]\right]\nonumber\\
& = \E\left[\sum_{h=1}^H  P_{k,h}^\top\left[ f^k_{\hat \theta_k}\left(U_{k,h}\right) - f^k_{\theta}\left( U_{k,h}\right)\right]\right]\nonumber\\
 &\leq \E\left[\sum_{h=1}^H \|P_{k,h}\|_2 \left\| f^k_{\hat \theta_k}\left(U_{k,h}\right) - f^k_{\theta}\left(U_{k,h}\right)\right\|_2\right]\nonumber\\
 & \leq qp_{\max}\E\left[\sum_{h=1}^H \left\| f^k_{\hat \theta_k}\left(U_{k,h}\right) - f^k_{\theta}\left(U_{k,h}\right)\right\|_2\right],\nonumber\
\end{align}
where the last inequality follows from the fact that the price vectors are in $[0, p_{\max}]^q$. 
\Halmos
\endproof

\subsubsection{Confidence Sets}
Define $W_{k,h}\in\R^q$ as
$$W_{k,h}[j] = \log Y_{k,h}[j] + \frac{\sigma^2}{2}, ~1\leq j\leq q.$$
Since $Y_{k,h}[j]$ is  log-normally distributed with parameters $D_{k,h}[j]-{\sigma^2}/{2}$ and $\sigma^2$, then $W_{k,h}$  is a random vector having a multivariate normal distribution of mean $D_{k,h}$ and covariance matrix $\sigma^2I_q$. 
 For any $k\in\N$ and given  the history $\HH_k$, let 
$$\bar\theta_k = \argmin_{\tilde\theta\in\Theta}{~\sum_{j=1}^{k-1}\sum_{h=1}^H \left\|W_{j,h} - f^k_{\tilde\theta}(P_{j,h},S_{j,h})\right\|_2^2}$$
be the least square estimate of the demand function parameter at the beginning of episode $k$. 

For given $\alpha>0$ and $\delta\in (0,1)$, let $N(\F,\alpha)$ denote the $\alpha$-covering number of $\F$ w.r.t. to the supremum norm and define for any $k\in \N$ 
$$\beta_k(\F,\delta,\alpha) = 8\sigma^2\log\left(N(\F,\alpha)/\delta\right) + 2\alpha kH\left(8d_{\max} + \sqrt{8\sigma^2\log(4k^2H^2/\delta)}\right).
$$
Similar to \cite{osband2014model}, we consider the following confidence set for $k\in\N$:
\begin{equation}
\label{confdef}
\C_k = \left\{\tilde \theta\in\Theta\Bigg|\sum_{j=1}^{k-1}\sum_{h=1}^H \left\|f^k_{\tilde\theta}(P_{j,h},S_{j,h}) - f^k_{\bar\theta}(P_{j,h},S_{j,h})\right\|_2^2\leq \beta_k(\F,\delta,\alpha) \right\}.
\end{equation}
The following proposition guarantees that  these confidence sets always contain the true demand function parameter with high probability.

\begin{proposition}[Prop. 5 of \cite{osband2014model}]
\label{prop1}
For any $\delta\in (0,1)$ and $\alpha>0$,
\begin{equation}
\label{conf2}
\P\left[\forall k\in\N:~\theta\in\C_k\right] \geq 1-2\delta.
\end{equation}
\end{proposition}

Given a particular $U = (P,S)$ and context vector $z$, we define the width of the confidence set $\C_k$ as 
\begin{equation}
\label{width}
w_k(U,z) = \sup_{\theta_1,\theta_2\in\C_k} \|f_{\theta_1}(U,z) - f_{\theta_2}(U,z)\|_2.
\end{equation}
The key element of our analysis is controlling the sum of the width of the  confidence sets defined in \eqref{confdef}. To do so, we present the following technical result from \cite{osband2014model}. 
\begin{lemma}[Prop. 6 of \cite{osband2014model}] 
\label{lem3}
Assume that $\|D\|_\infty\leq d_{\max}$ always hold. For any $K\in\N$ and any sequence  $(z_1,U_{1,1},\cdots,U_{1,H},\cdots,z_K,U_{K,1},\cdots,U_{K,H})$, we have
\begin{equation}
\label{lem3eq}
\sum_{k=1}^K\sum_{h=1}^H w_k(U_{k,h},z_k) \leq 1 + Hd_{\max} d_E(\F,(KH)^{-1}) + 4\sqrt{\beta_Kd_E(\F,(KH)^{-1})KH}.
\end{equation}
\end{lemma}

\subsubsection{Proof of Theorem \ref{mainth0}}
Now, we have all the necessary tools to prove Theorem \ref{mainth0}.

\proof{Proof of Theorem \ref{mainth0}.}
First note that for any $k\in N$ and given  history $\HH_k$ at the start of episode $k$, the sampled parameter $\hat\theta_k$ is identically distributed with the true parameter $\theta$. Therefore from Proposition \ref{prop1}, it follows that $\P\left[\forall k\in\N:~\hat\theta_k\in\C_k\right] \geq 1-2\delta$. Now, define the event $\mathcal{E}_k = \{\theta,\hat\theta_k\in\C_k\}$, and let $\mathcal{E}_k^c$ be its complement.
By Lemma \ref{lem15}, we have 
\begin{align*}
R_K(TP) & = \sum_{k=1}^K \Delta_k\\
& \leq qp_{\max}\sum_{k=1}^K\E\left[\sum_{h=1}^H \left\| f^k_{\hat \theta_k}\left(U_{k,h}\right) - f^k_{\theta}\left(U_{k,h}\right)\right\|_2\right]\\
& = qp_{\max} \E\left[\sum_{k=1}^K \sum_{h=1}^H \left\| f^k_{\hat \theta_k}\left(U_{k,h}\right) - f^k_{\theta}\left(U_{k,h}\right)\right\|_2\1(\mathcal{E}_k)\right]\\
&\hspace{7em}+qp_{\max} \E\left[\sum_{k=1}^K \sum_{h=1}^H \left\| f^k_{\hat \theta_k}\left(U_{k,h}\right) - f^k_{\theta}\left(U_{k,h}\right)\right\|_2\1(\mathcal{E}_k^c)\right]\\
&\leq qp_{\max} \E\left[\sum_{k=1}^K \sum_{h=1}^Hw_k(U_{k,h},z_k)\right] + 2qp_{\max}d_{\max}\E\left[\sum_{k=1}^K \sum_{h=1}^H \1(\mathcal{E}_k^c)\right]\\
&\leq qp_{\max} \E\left[\sum_{k=1}^K \sum_{h=1}^Hw_k(U_{k,h},z_k)\right] + 2qp_{\max}d_{\max}KH\P\left[\theta\notin\cap_{i\in\N}\C_i \mbox{ or }\hat\theta_k\notin\cap_{i\in\N}\C_i\right]\\
&\stackrel{(a)}{\leq} qp_{\max}\E\left[\sum_{k=1}^K \sum_{h=1}^Hw_k(U_{k,h},z_k)\right] + 4qp_{\max}d_{\max}KH\delta\\
& \stackrel{(b)}{\leq} qp_{\max}\left(1 + Hd_{\max} d_E(\F,(KH)^{-1}) + 4\sqrt{\beta_Kd_E(\F,(KH)^{-1})KH}\right) + 4qp_{\max}d_{\max}KH\delta,
\end{align*}
where $(a)$ and $(b)$ follow from Proposition \ref{prop1} and Lemma \ref{lem3}, respectively. Now  we take $\delta = \alpha = (KH)^{-2}$, which gives 
$$
\beta_K =8\sigma^2\log((KH)^2 N(\F,(KH)^{-2}))+\frac{2}{KH} \left(8d_{\max} + \sqrt{8\sigma^2\log 4}\right),$$
and
\begin{equation}
\label{mainpfeq1}
R_K(TP)\leq qp_{\max}\left(1 + Hd_{\max} d_E(\F,(KH)^{-1}) + 4\sqrt{\beta_Kd_E(\F,(KH)^{-1})KH}\right) + \frac{4qp_{\max}d_{\max}}{KH}.
\end{equation}
Also, as has been shown in Proposition 7 of \cite{dan1}, we have the following relation between $\beta_k$ and  Kolmogorov dimension of $\F$:
$$\beta_K(\F,(KH)^{-2},(KH)^{-2}) = 16\sigma^2(1 + o(1) + d_K(\F)) \log(KH),$$
which gives $\beta_K = O(\sigma^2 d_K(\F)\log (HK))$. 
Combining this with \eqref{mainpfeq1} gives the second  statement of the theorem.
\Halmos
\endproof

\subsection{Other Technical Results}
Theorem \ref{mainth0} provides a regret bound for TP in a general scenario involving multiproduct pricing, covariates and reference effects. As a result of this generality, all other theoretical results of this paper can be derived using Theorem \ref{mainth0}.  For example a proof for Theorem \ref{mainth} is as follows. 

\proof{Proof of Theorem \ref{mainth}.}
The setting considered in Theorem \ref{mainth0} can be thought of a special case of the setting of Theorem \ref{mainth0} where $q=1$ and all covariate vectors are the singleton $\forall~k:~z_k=1$. Thus, the statement of Theorem \ref{mainth} follows immediately from Theorem \ref{mainth0} by setting $q=1$ and $\Z=\{1\}$ (i.e., the covariates effect can be neglected). 
\Halmos
\endproof

The following is a proof for Theorem \ref{reglinstcov}. 
\proof{Proof of Theorem \ref{reglinstcov}.}
The settin of Theorem \ref{reglinstcov} can be thought of as a special case of that of Theorem \ref{mainth0} where the number of products being sold at any episode is $q=1$ and the demand function $f_\theta$ is a linear function as in \eqref{expdemstlincov}. Let $\F$ be the class of such demand functions spanned by the parameter $\theta$. It is easy to see that under Assumptions \ref{assump2} and \ref{assump3}, we have (see, for example, Proposition 2 of \cite{osband2014model})
$$d_K(\F) = O(mn^2),~~~~d_E(\F,\epsilon) = O(mn^2\log(np_{\max}\lambda\tau/\epsilon).$$
The statement then follows from Theorem \ref{mainth0}.
\Halmos
\endproof

The following is a proof of Theorem \ref{multtheorem}. 
\proof{Proof of Theorem \ref{multtheorem}.}
The setting of Theorem \ref{multtheorem} can be thought of as a special case of that of Theorem \ref{mainth0} with $\Z=\{1\}$ such that the covariates do not matter. Then, with $\F$ as the class of demand function as defined in \eqref{expdemstlincovmulti}, it is easy to see that (see, for example, Proposition 2 of \cite{osband2014model}) under Assumptions \ref{assump2} and  \ref{assump4}
$$d_K(\F) = O(q^2n^2),~~~~~d_E(\F,\epsilon) = O(q^2n^2\log(nqp_{\max}\tau/\epsilon)).$$
the statement then follows from Theorem \ref{mainth0}.
\Halmos
\endproof

\section{Concluding Remarks}
\label{ConcSec}

We studied the problem of dynamic pricing in an unknown  environment in the presence of reference effects.  Our framework accommodates contexts in which consumers make purchase decisions based on price histories, not only prevailing prices.  The fact that prices impose delayed consequences introduces challenges that call for more sophisticated pricing strategies.  In particular, the seller can learn to influence consumer behavior by judiciously sequencing prices.

To address this challenge, we formulated the dynamic pricing problem in terms of reinforcement learning.  In our framework, the demand for an item depends on its current price, price history, and unknown parameters of a demand model.  We allow for arbitrary demand functions and propose Thompson Pricing (TP) as a heuristic for addressing the problem. 
We provided a very general regret bound on the performance of TP in terms of the eluder dimension and Kolmogorov dimensions of the demand function class. 

We also presented extensions of TP that address contexts with observable demand covariates, multiproduct pricing, and varying and overlapping sales cycles. We provided a general performance analysis of the TP, which specializes to offer regret bounds for each of the aforementioned scenarios.

\ACKNOWLEDGMENT{The authors would like to thank Mohsen Bayati for his constructive feedback and discussion on this paper. This work was generously supported by  a Stanford Graduate Fellowship.
}


\end{document}